# Identifying and Supporting Financially Vulnerable Consumers in a Privacy-Preserving Manner: A Use Case Using Decentralised Identifiers and Verifiable Credentials

Tasos Spiliotopoulos[1], Dave Horsfall[2], Magdalene Ng[2], Kovila Coopamootoo[2], Aad van Moorsel[2], and Karen Elliott[1]

[1]Business School, Newcastle University, United Kingdom, {firstname.lastname}@newcastle.ac.uk

[2]School of Computing, Newcastle University, United Kingdom, {firstname.lastname}@newcastle.ac.uk

Vulnerable individuals have a limited ability to make reasonable financial decisions and choices and, thus, the level of care that is appropriate to be provided to them by financial institutions may be different from that required for other consumers. Therefore, identifying vulnerability is of central importance for the design and effective provision of financial services and products. However, validating the information that customers share and respecting their privacy are both particularly important in finance and this poses a challenge for identifying and caring for vulnerable populations. This position paper examines the potential of the combination of two emerging technologies, Decentralized Identifiers (DIDs) and Verifiable Credentials (VCs), for the identification of vulnerable consumers in finance in an efficient and privacy-preserving manner.

CCS CONCEPTS • Human-centered computing → Human computer interaction (HCI) → HCI theory, concepts and models • Security and privacy → Cryptography • Social and professional topics → User characteristics • Information systems → Information systems applications

**Additional Keywords and Phrases:** vulnerability, finance, self-sovereign identities (SSIs), decentralised identifiers (DIDs), verifiable credentials (VCs), selective disclosure, privacy, security

## 1 INTRODUCTION

The identification of vulnerable consumers is of great importance in the financial sector, because it allows financial institutions to train staff, allocate resources and design products and services in a way that supports vulnerable populations. For this reason, the financial regulatory body in the United Kingdom, the Financial Conduct Authority (FCA), has been tracking the vulnerability of consumers over time, together with other aspects of their financial lives. A large-scale tracking survey by the regulator has found that, before Covid-19, the number of UK adults showing one or more characteristics of vulnerability was decreasing, with this decrease largely attributed to improvements in digital inclusion and financial resilience. However, the latest results of this survey show that Covid-19 has reversed this positive trend in vulnerability and has disproportionately affected specific population groups, such as younger adults and the self-employed [4]. Besides urgency, this finding highlights the need for more nuance and flexibility towards the identification and provision of appropriate care to vulnerable populations. This position paper provides an overview of financial vulnerability and discusses the use of technologies for the management of digital identities to address this need. In particular, we examine the potential of Decentralized Identifiers (DIDs) and Verifiable Credentials (VCs), two emerging standards from the World Wide Web Consortium (W3C) that can be combined to provide a decentralized approach for reliable identity assurance, authentication, and attestations. Key characteristics of this approach are *self-sovereignty*, i.e., that people and businesses store and control their data on their own devices and provide these data only when someone needs to validate them, and *selective disclosure*, i.e., that only relevant private information is shared with interested parties in a privacy-preserving way.

## 2 DIGITAL IDENTIFIERS AND VERIFIABLE CREDENTIALS

DIDs constitute a new type of identifier that is i) *decentralised*, i.e., there is no central issuing agency; ii) *persistent*, i.e., does not require the continued operation of an underling organization; iii) *cryptographically verifiable*, i.e., it is possible to prove control of the identifier cryptographically; and iv) *resolvable*, i.e., it is possible to discover metadata about the identifier [9]. This means that individuals and organizations are able to generate their own globally unique identifiers using systems they trust, and to prove control of those identifiers (i.e., authenticate themselves) using cryptographic proofs, such as digital signatures.

VCs are the electronic equivalent of the physical credentials we all possess today, such as passports, plastic cards, driving licences, qualifications and awards. A VC can represent all of the same information that a physical credential represents, including information related to identifying the subject of the credential (e.g., photograph, name), information related to the issuing authority (e.g., a city government, certification body), information related to constraints on the credential (e.g., expiration date) and so on. Importantly, the issuer and the subject of a VC can be identified by DIDs and digitally signed, thus making VCs tamper-evident and more trustworthy than their physical counterparts. So, as VCs are tamper-evident and their authorship (i.e., the issuer of the credential) can be cryptographically verified, the claims included in a VC (such as the fact that someone is a UK citizen or has a degree from a specific university), can be fully trusted as long as the issuer is trustworthy [10]. Furthermore, VCs can be stored in a credential repository within a user agent (e.g., a mobile app) managed by the credential holder, so that the holder has complete control over who is authorised to access a VC and the holder can even share a revocable or time-limited token with a verifier that allows them to check the status of claims over time.

One or more VCs can be used to build a *verifiable presentation*, which allows a holder to share data with a verifier in such a way that the authorship of the data is still cryptographically verified [10]. A verifiable presentation can also be shared by a holder without revealing the identity of the verifier to the issuer. Crucially, verifiable presentations (as assembled by holders) and VCs (as issued by issuers) can both support selective disclosure [7,10]. This means that holders can present proofs of claims in a VC without revealing the entire VC. This may be through the ability of a holder to select some elements of a verifiable credential to share with a verifier, without revealing the rest. Alternatively, the presentation may satisfy some derived predicates requested by the verifier. These predicates can be expressed as Boolean conditions such as greater than, less than, equal to, is in set etc. For example, someone can prove that they are over 18 years of age by selectively disclosing information based on the appropriate predicate (i.e., 'age > 18') from their passport (their VC), without disclosing any other information in their passport, not even their exact age. Other examples include proving that someone maintains a legal disability status (e.g., based on a VC from the National Health Service) without disclosing the specific disability, or proving that someone has income above a certain level (e.g., based on a VC issued by their bank) in order to be eligible to rent an apartment, without disclosing their exact income or other assets. This ability to selectively disclose personal information is particularly important for vulnerable individuals, as their source of vulnerability may be something that they would not like to disclose to a financial institution in detail or the institution may not feel comfortable requesting on their own.

## 3  IDENTIFYING VULNERABLE CONSUMERS IN FINANCE

The FCA has identified a vulnerable consumer as 'somebody who, due to their personal circumstances, is especially susceptible to harm, particularly when a firm is not acting with appropriate levels of care' [3]. The UK financial regulator has identified the protection of vulnerable customers as a key priority for the industry and has published appropriate guidance for financial firms, strongly encouraging them to treat vulnerable customers fairly [3]. Four categories of characteristics are considered to constitute drivers of financial vulnerability – poor health, impact of life events, low resilience and low capability [3] (Table 1) - with the latest report finding that 53% of UK adults show one or more of these characteristics [4].

In terms of these vulnerability characteristics, Covid-19 has had a substantial impact in the 'life events' category, in particular causing redundancy or reduced working hours, and the 'resilience' category. Low resilience is described as being over-indebted or having little capacity to withstand financial shocks, such as losing the main source of household income for even a week [3]. Covid-19 was also found to have a negative impact on mental health [4,5], something that can result in a range of difficulties when dealing with financial services. Although not explicitly identified by the FCA, other work has established effects of Covid-19 on other vulnerability characteristics, such as domestic abuse [13] and bereavement [12].

People with vulnerability characteristics are more likely to lack confidence in the financial industry, and this lack of trust and confidence can result in consumers not engaging with financial services and products, or failing to address their own financial needs [4]. In order to address this problem and support vulnerable individuals more effectively, financial firms are encouraged to i) understand vulnerable customers by carrying out research and collecting vulnerability data, ii) train and develop their staff to embed the consideration of vulnerable consumers, iii) consider the communication and information needs of vulnerable customers, iv) adapt customer service processes and systems to account for vulnerable customers, v) integrate the consideration of vulnerable customers into the product and service design process, and vi) monitor and evaluate the treatment of vulnerable customers [3].

Table 1. The four key drivers of vulnerability and the types of characteristics of vulnerability they may cause (adapted from FCA guidance [3])

| Health | Life events | Resilience | Capability |
|---|---|---|---|
| Physical disability | Caring responsibilities | Low or erratic income | Low knowledge or confidence in managing finances |
| Severe or long-term illness | Bereavement | Over indebtedness | Poor literacy or numeracy skills |
| Hearing or visual impairment | Income shock | Low savings | Low English language skills |
| Poor mental health | Relationship breakdown | Low emotional resilience | Poor or non-existent digital skills |
| Addiction | Domestic abuse | | Learning impairments |
| Low mental capacity or cognitive impairment | People with non-standard requirements such as people with convictions, care leavers, refugees | | No or low access to help or support |
| | Retirement | | |

## 4 DISCUSSION

While there is a range of diverse vulnerability characteristics than can affect the financial lives of individuals, here we discuss an overarching scenario that can constitute an example or the starting point for more complex use cases. The scenario involves a customer that collects VCs through the day-to-day interaction with other parties. These credentials are stored in a credential repository in the user agent (i.e., an app) installed in the user's smartphone. Some of these credentials may describe a type of vulnerability, for example after a visit to the hospital the user may receive a VC that can be used to prove a permanent or temporary physical disability. A financial service provider (e.g., a bank) during a regular or ad hoc vulnerability check, may request information from a VC that can be used to make a claim about vulnerability. Alternatively, the customer may issue the bank with a revocable token that allows the bank to periodically (e.g., every two weeks) check the current vulnerability status of the consumer. The bank assigns a 'vulnerability flag' to the customer based on specific guidelines and makes efforts to provide appropriate care (e.g., payment holidays or a pause on interest and fees).

The self-sovereignty and selective disclosure that are afforded by this combination of DIDs and VCs not only can empower vulnerable customers in their interaction with financial institutions, but also provide benefits to other entities involved. In the above example, the VC issuer (i.e., the hospital) doesn't bear the risk of storing sensitive personal data; the verifier (i.e., the bank) can be certain of the authenticity of any claims; and the VC holder (i.e., the customer) has control of any information and can revoke access to it at any time. Furthermore, a possible periodical, automatic check of vulnerability can help identify financial problems before they lead to debt spirals. Such a periodic check can be considered especially pertinent lately, as the Covid-19 pandemic has substantially changed the consumer mindset and behaviour, making historical data on customers less valid than they used to be [11].

From a privacy perspective, the customer has control over the disclosure of the information held in the VC and can engage in selective disclosure in a privacy-preserving way – in this case they receive a 'vulnerability flag' without disclosing the details of their specific disability. Furthermore, a revocable token aligns such an implementation with the consumers' 'right to be forgotten' [8]. While HCI research is starting to recognise the value of privacy for vulnerable populations in other contexts [6], this privacy-by-design approach that can be achieved through this selective disclosure can help close the gap between trustworthiness and trust in financial services [1]. That is, as users know that a system is designed with their privacy in mind and that other stakeholders cannot breach this privacy, they are bound to trust this interaction more. Still, while the bank does not have access to the details of the vulnerabilities for each individual customer, it has access to aggregate data that allow it to train staff, allocate resources, and design products and services in a way that supports vulnerable consumers.

Of course, a simple 'vulnerability flag' can be too coarse to describe the details, nuances, and diverse types of vulnerability that may need to be identified in order to provide appropriate care to the affected individuals. The FCA guidance describes an algorithm for calculating vulnerability based on their survey questions [4]. While such an algorithm is admittedly incomplete, it can still be a starting point for the design of VCs and their interaction with verifiers with the goal of getting insights into vulnerability measurement in this context. From a technical perspective, questions about where and how a 'vulnerability calculation' takes place and what precise data are required can only be addressed on the basis of specific use cases. The next steps in our work involve a systematic mapping of vulnerability characteristics to specific VC implementations and use cases informed by engagement with stakeholders from the financial sector, and a pilot deployment of the proposed DID and VC implementation in a sandbox environment that can be evaluated with end users and other stakeholders.

Although DIDs and VCs are still emerging specifications, a growing body of research is starting to identify the potential of technologies based on this kind of identity management across diverse sectors (e.g., [2,14]). This position paper has focused on the use case of identifying financially vulnerable consumers and presented a framework that can potentially be adapted to identify financial vulnerabilities in other countries and vulnerabilities in different contexts.

## ACKNOWLEDGMENTS

This work was supported, in whole or in part, by the Bill & Melinda Gates Foundation [INV-001309]. Under the grant conditions of the Foundation, a Creative Commons Attribution 4.0 Generic License has already been assigned to the Author Accepted Manuscript version that might arise from this submission.